\documentclass[a4paper,12pt, epsfig]{article}
\def\letter{0}\def\pr{0}
\pdfoutput=1 
\usepackage{epsfig}
\usepackage{epstopdf}
\usepackage{textcomp}
\usepackage{graphicx}
\usepackage{ifthen}
\usepackage{ulem}

\usepackage{amsmath}
\usepackage[psamsfonts]{amssymb}
\usepackage{euscript}

\usepackage{latexsym}
\usepackage[arrow,matrix,curve]{xy}

\usepackage[hypertexnames=false]{hyperref}
\hypersetup{
    colorlinks=true,
    linkcolor=blue,
    filecolor=magenta,   
    citecolor=black,   
    urlcolor=cyan,
    }

\newskip\humongous \humongous=0pt plus 1000pt minus 1000pt

\newif\ifdtup

\def\,{\hspace{-.1cm}}
\def\hsp{,\hspace{.7cm}}

\def\ddp#1{d^3\vp_{#1}}
\def\dvx{\int d^3\vx}

\def\ddpp#1{d^3\vpp_{#1}}

\def\fc#1#2 {\frac{n}{q}#1\frac{n}{q}#2}

\def\kt{\kappa}
\def\kt{\mathfrak{K}}
\def\ks{|\kt\rangle}

\def\hp{H^{\prime(t)}}
\def\rv{{\rm{vac}}}

\newcommand{\vac}{\ensuremath{|0\rangle}}
\newcommand{\ovac}{\ensuremath{|\Omega\rangle}}
\newcommand{\ps}{\ensuremath{|\vp_0\rangle}}

\renewcommand{\sin}{\textrm{sin}}

\renewcommand{\sinh}{\textrm{sinh}}
\renewcommand{\cosh}{\textrm{cosh}}
\renewcommand{\tanh}{\textrm{tanh}}
\newcommand{\sech}{\textrm{sech}}
\newcommand{\csch}{\textrm{csch}}

\def\exp#1{\hbox{\rm exp}\left[#1\right]}

\def\sign#1{{\rm sign}\left(#1\right)}

\renewcommand{\theequation}{\arabic{section}.\arabic{equation}}
\renewcommand{\(}{\begin{equation}}
\renewcommand{\)}{end{equation} \vspace{-.05in}\linebreak}

\newcounter{saveeqn}
\newcounter{savealpheqn}

\newcommand{\alpheqn}{\setcounter{saveeqn}{\value{equation}}%
  \stepcounter{saveeqn}\setcounter{equation}{0}%
  \renewcommand{\theequation}{\mbox{\arabic{section}.\arabic{saveeqn}
\alph{equation}}}
  \renewcommand{\)}{\end{equation}}}
\def\part#1{\frac{\partial}{\partial{#1}}}%
\def\group#1{\refstepcounter{equation}\setcounter{saveeqn}
 {\value{equation}}%
  \label{#1}\setcounter{equation}{0}%
\renewcommand{\theequation}{\mbox{\arabic{section}.\arabic{saveeqn}
\alph{equation}}}
  \renewcommand{\)}{\end{equation}}}
\newcommand{\reseteqn}{\setcounter{equation}{\value{saveeqn}}%
  \renewcommand{\theequation}{\arabic{section}.\arabic{equation}}%
  \renewcommand{\)}{\end{equation}}}

\newcommand{\aalpheqn}{\setcounter{saveeqn}{\value{equation}}%
  \stepcounter{saveeqn}\setcounter{equation}{0}%
  \renewcommand{\theequation}{\mbox{
        \Alph{subsection}.\arabic{saveeqn}\alph{equation}}}
   \renewcommand{\)}{\end{equation}}}
\newcommand{\areseteqn}{\setcounter{equation}{\value{saveeqn}}%
  \renewcommand{\theequation}{\Alph{subsection}.\arabic{equation}}%
  \renewcommand{\)}{\end{equation}}}

\renewcommand{\thefootnote}{\alph{footnote}}
\renewcommand{\(}{\begin{equation}}
\renewcommand{\)}{\end{equation}}
\newcommand{\ba}{\begin{eqnarray}}
\newcommand{\ea}{\end{eqnarray}}

\renewcommand{\b}{\beta}

\renewcommand{\sl}{{\sqrt{\lambda}}}

\newcommand{\cbp}{\mathop{\vtop{\ialign{##\crcr
   $\hfil\displaystyle{}\hfil$\crcr\noalign{\kern-13pt\nointerlineskip}
   \BIG{)}\hskip0pt\crcr\noalign{\kern3pt}}}}}
\newcommand{\pa}{\mathop{\vtop{\ialign{##\crcr

$\hfil\displaystyle{\oplus}\hfil$\crcr\noalign{\kern+1pt\nointerlineskip
}
   \hspace{.08in}$^{\alpha=0}$\hskip6pt\crcr\noalign{\kern3pt}}}}}
\renewcommand{\hsp}{,\hspace{.3in}}
\newcommand{\p}{^\prime}
\newcommand{\pp}{^{\prime\prime}}

\catcode`\@=11
\def\vereq#1#2{\lower3pt\vbox{\baselineskip1.5pt \lineskip1.5pt
\ialign{$\m@th#1\hfill##\hfil$\crcr#2\crcr\sim\crcr}}}
\catcode`\@=12

\renewcommand{\(}{\begin{equation}}
\renewcommand{\)}{\end{equation}}

\def\vx{{\vec{x}}}
\def\vp{{\vec{p}}}
\def\vpp{{\vec{p}^{
\hspace{.05cm}\prime
}}}
\def\vppp{{\vec{p}^{
\hspace{.05cm}\prime\prime
}}}
\def\vk{{\vec{k}}}
\def\vkp{{\vec{k}\p}}

\def\pin#1{\int \frac{d#1}{2\pi}}
\def\ppin#1{\int\hspace{-17pt}\sum \frac{d#1}{2\pi}}
\def\ppink#1{\int\hspace{-17pt}\sum\frac{d^{#1}\vk}{(2\pi)^{#1}}}

\def\ppinkp#1{\int\hspace{-17pt}\sum\frac{d^{#1}k\p}{(2\pi)^{#1}}}
\def\dint{\int\hspace{-12pt}\sum }
\def\pink#1{\int \frac{d^{#1}k}{(2\pi)^{#1}}}

\def\pinkp#1{\int \frac{d^{#1}k\p}{(2\pi)^{#1}}}

\def\pinvp#1{\int \frac{d^{#1}\vp}{(2\pi)^{#1}}}
\def\pinvk#1{\int\hspace{-17pt}\sum \frac{d^{#1}\vk}{(2\pi)^{#1}}}
\def\kinv#1#2{\int\hspace{-17pt}\sum \frac{d^{#1}\vec{#2}}{(2\pi)^{#1}}}
\def\pinv#1#2{\int \frac{d^{#1}\vec{#2}}{(2\pi)^{#1}}}
\def\pinvx#1#2{\int d^{#1}\vec{#2}}
\def\pinvi#1#2#3{\int \frac{d^{#1}\vec{#2}_{#3}}{(2\pi)^{#1}}}
\def\ppinv#1#2{\int \frac{d^{#1}\vec{#2}^{
\hspace{.07cm}\prime
}}{(2\pi)^{#1}}}

\def\Bd#1{B^\ddag_{k_{#1}}}
\def\Ad#1{A^\ddag_{\vp_{#1}}}
\def\Btd#1{B^{(t)\ddag}_{k_{#1}}}
\def\Bt#1{B^{(t)}_{k_{#1}}}
\def\Bdp#1{B^\ddag_{k\p_{#1}}}

\def\cc{\mathcal{C}}
\def\df{\mathcal{D}_{f}}
\def\dv{\mathcal{D}_{v}}
\def\dft{\mathcal{D}_{f}^{(t)}}
\def\dfd{\mathcal{D}_{f}^{(t)\dag}}

\def\avkd{A^\ddag_{\vec{k}}}
\def\avpd{A^\ddag_{\vec{p}}}
\def\avkm{A_{-\vec{k}}}
\def\avpm{A_{-\vec{p}}}
\def\avk{A_{\vec{k}}}
\def\avp{A_{\vec{p}}}
\def\omvk{\omega_{\vec{k}}}
\def\omvp{\omega_{\vec{p}}}

\def\navkd#1{A^\ddag_{\vec{k}_{#1}}}
\def\navpd#1{A^\ddag_{\vec{p}_{#1}}}

\def\navpm#1{A_{-\vec{p}_{#1}}}
\def\navk#1{A_{\vec{k}_{#1}}}
\def\navp#1{A_{\vec{p}_{#1}}}

\def\nomvp#1{\omega_{\vec{p}_{#1}}}

\def\O#1#2{|\Omega\rangle_{#1}^{#2}}

\def\I{\mathcal{I}}

\def\os{\omega_S}

\def\gx{(\gamma(x-vt))}

\def\jhep#1{\textcolor{green}{#1}}
\def\red#1{\textcolor{red}{Jarah: #1}}
\def\blu#1{\textcolor{blue}{Hui: #1}}

\usepackage[dvipsnames]{xcolor}
\def\gre#1{\textcolor{ForestGreen}{Hengyuan: #1}}
\def\hui#1{\textcolor{Mulberry}{Hui: #1}}

\newcommand{\beas}{\begin{eqnarray*}}
\newcommand{\eeas}{\end{eqnarray*}}

\newcommand{\bquo}{\begin{quote}}
\newcommand{\enqu}{\end{quote}}

\def\lim#1{\stackrel{\rm{lim}}{{}_{#1}}}

\newcommand{\R}{{\mathbb R}}

    \newcommand{\g}{\mathfrak g}

\def\ch{{\mathcal{H}}}
\def\co{{\mathcal{O}}}

\def\ok#1{\omega_{k_{#1}}}
\def\okp#1{\omega_{k\p_{#1}}}
\def\ovp#1{\omega_{\vp_{#1}}}
\def\ovpp#1{\omega_{\vpp_{#1}}}
\def\ovppp#1{\omega_{\vppp_{#1}}}
\def\okt#1{\omega_{\kt_{#1}}}
\def\V#1{V^{(#1)}(\sqrt{\lambda}f(x))}
\def\Vg#1{V^{(#1)}(\sqrt{\lambda}f(\gamma(x-vt))}
\def\ck{\csch\left(\frac{\pi k}{2\b}\right)}

\def\mb{\mathcal{B}}
\def\mc{\mathcal{C}}
\def\md{\mathcal{D}}
\def\me{\mathcal{E}}
\def\gt{\tilde{\g}}
\def\dim{2}
\def\dimtwo{4}
\def\dimthree{6}
\def\phip{\phi}
\def\pip{\pi}

\newcommand{\beq}{\begin{equation}}
\newcommand{\eeq}{\end{equation}}
\newcommand{\bea}{\begin{eqnarray}}
\newcommand{\eea}{\end{eqnarray}}

\newskip\humongous \humongous=0pt plus 1000pt minus 1000pt

\newif\ifdtup

\catcode`\@=11

\ifthenelse{\equal{\letter}{0}}{ 

\@addtoreset{equation}{section}
\def\theequation{\arabic{section}.\arabic{equation}}

\def\@normalsize{\@setsize\normalsize{15pt}\xiipt\@xiipt
\abovedisplayskip 14pt plus3pt minus3pt%
\belowdisplayskip \abovedisplayskip
\abovedisplayshortskip \z@ plus3pt%
\belowdisplayshortskip 7pt plus3.5pt minus0pt}

\def\small{\@setsize\small{13.6pt}\xipt\@xipt
\abovedisplayskip 13pt plus3pt minus3pt%
\belowdisplayskip \abovedisplayskip
\abovedisplayshortskip \z@ plus3pt%
\belowdisplayshortskip 7pt plus3.5pt minus0pt
\def\@listi{\parsep 4.5pt plus 2pt minus 1pt
      \itemsep \parsep
      \topsep 9pt plus 3pt minus 3pt}}

\relax

\def\section{\@startsection{section}{1}{\z@}{3.5ex plus 1ex minus  .2ex}{2.3ex plus .2ex}{\large\bf}}

\def\thesection{\arabic{section}}
\def\thesubsection{\arabic{section}.\arabic{subsection}}

\def\appendix{\setcounter{section}{0}
 \def\thesection{Appendix \Alph{section}}
 \def\thesubsection{\Alph{section}.\arabic{subsection}}
 \def\theequation{\Alph{section}.\arabic{equation}}}
\renewcommand{\theequation}{\arabic{section}.\arabic{equation}}

}{
\renewcommand{\theequation}{\arabic{equation}}

} 

\begin{document}
\def\thefootnote{\fnsymbol{footnote}}
\def\thetitle{Scheme Dependence of the One-Loop Domain Wall Tension}
\def\autone{Jarah Evslin}
\def\autthree{Hui Liu}
\def\autfour{Baiyang Zhang}
\def\auttwo{Hengyuan Guo}
\def\affd{Yerevan Physics Institute, 2 Alikhanyan Brothers St., Yerevan, 0036, Armenia}
\def\affb{University of the Chinese Academy of Sciences, YuQuanLu 19A, Beijing 100049, China}
\def\affa{Institute of Modern Physics, NanChangLu 509, Lanzhou 730000, China}
\def\affc{School of Physics and Astronomy, Sun Yat-sen University, Zhuhai 519082, China}
\def\affe{Institute of Contemporary Mathematics, School of Mathematics and Statistics, Henan University, Kaifeng, Henan 475004, P. R. China}


\ifthenelse{\equal{\pr}{1}}{
\title{\thetitle}
\author{\autone}
\author{\auttwo}
\author{\autthree}
\affiliation {\affa}
\affiliation {\affb}
\affiliation {\affc}

}{

\begin{center}
{\large {\bf \thetitle}}

\bigskip

\bigskip



{\large \noindent  \autone{${}^{1,2}$} \footnote{jarah@impcas.ac.cn} and \autthree{${}^{3}$} \footnote{hui.liu@yerphi.am}}

\vskip.7cm

1) \affa\\
2) \affb\\
3) \affd\\

\end{center}

}

\begin{abstract}
\noindent
The one-loop tension of the domain wall in the 3+1 dimensional $\phi^4$ double-well model was derived long ago using dimensional regularization.  The methods used can only be applied to solitons depending on a single dimension.  In the past few months,  domain wall tensions have been recalculated using spectral methods with Born subtractions and also linearized soliton perturbation theory, both of which may be generalized to arbitrary solitons.  It has been shown that the former agrees with the results of Rebhan et al.  In the present work, we argue that, if the same renormalization scheme is chosen, both new results agree. 

\end{abstract}

\section{Introduction}

The one-loop quantum correction to the tension of the kink in the 1+1 dimensional $\phi^4$ double-well model was computed half a century ago in Ref.~\cite{dhn2}.  It was not long before the result was generalized to the domain wall in the corresponding 3+1 dimensional model, in Ref.~\cite{bf84}.  These results were confirmed in later studies \cite{munster89,rebmuro}.

However, there was reason for concern.  First of all, some studies found incompatible answers \cite{car1,car2}.  More importantly, in Ref.~\cite{rebhan} it was realized that conventional regularization schemes, such as a sharp cutoff, can yield incorrect quantum mass corrections.  This surprising result led to the suggestion in Ref.~\cite{lit} that regularization schemes be fully consistent with the deformations brought to the soliton by the regularization itself.  In other words, in the regularized theory the soliton solution and its normal modes are different, and one can only trust a regularization if they treated consistently for each value of the regulator.  Needless to say, a noninteger dimension, even if only parallel to the soliton, corresponds to an unphysical situation in which no consistent soliton solution can be defined.  This motivates a new calculation without dimensional regularization.

The dimensional regularization approaches, as stressed in Ref.~\cite{py00}, suffer from another limitation.  They can only be applied to solitons that depend nontrivially on a single dimension, in other words only to domain walls.  However many other solitons, such as monopoles, are of phenomenological interest.

Spurred on by recent interest in cosmology \cite{bp21,bp23}, this last shortcoming was treated recently in Ref.~\cite{gw}, albeit within the context of dimensional regularization.  Satisfyingly, its results are in complete agreement with Ref.~\cite{rebmuro}.  Similarly, recently the tension was calculated using linearized soliton perturbation theory in Ref.~\cite{noifin}.  This approach is based on the Hamiltonian methods of Ref.~\cite{cahill76} and so is compatible with normal ordering, which removes the polynomial ultraviolet divergences allowing one to use a simple cutoff, obviating the need for dimensional regularization.

In the present note, we show that Refs.~\cite{gw} and \cite{noifin} are consistent, with only a minor difference in the choice of renormalization scheme.  Namely, while both use a scheme in which quantum corrections to the three-point coupling vanish, the three-point coupling itself is defined in different vacua in the two studies.  We provide a simple, analytical formula for the change in the one-loop tension correction resulting from this change in scheme, which agrees with the mismatch in the results of these references in all dimensions considered. 

We begin in 1+1 dimensions in Sec.~\ref{ccg}, reviewing the one-loop quantum correction to the kink mass calculated in Ref.~\cite{cahill76}.  This will be more convenient than the approach of Ref.~\cite{dhn2} because it uses a normal-ordered Hamiltonian, which removes all ultraviolet divergences in 1+1 dimensions and, at one loop, even in 2+1 dimensions.  In Sec.~\ref{arbsez} we consider an arbitrary renormalization scheme, characterized by a renormalization of the meson mass and coupling.  In this context, we define a one-loop mass shift.  In Sec.~\ref{rensez} we calculate the potentially-divergent contributions to the amputated two-point and three-point functions.  Finally, in Sec.~\ref{tensez}, we consider two renormalization schemes, which tie together the loop diagrams of Sec.~\ref{rensez} with the arbitrary renormalization of Sec.~\ref{arbsez}.  The first scheme leads to the results of Ref.~\cite{noifin} and the second to those of Ref.~\cite{gw}.  We then lift the argument from 1+1 to 2+1 and 3+1 dimensions, which turns out to be rather trivial and reproduces, analytically, the differences reported between Refs.~\cite{gw} and \cite{noifin}.

\section{Cahill, Comtets and Glauber} \label{ccg}

Consider the $\phi^4$ double-well model in 1+1 dimensions.  The Hamiltonian is
\beq
H=\int dx\left[\frac{:\pi^2:+:\left(\partial_x\phi\right)^2:}{2}+\frac{\lambda_0}{4}:\left(\phi^2-\frac{m_0^2}{2\lambda_0}\right)^2:\right] \label{heq}
\eeq
where $::$ is the usual Schrodinger picture normal ordering in terms of a plane wave decomposition, and $\phi(x)$ and $\pi(x)$ are a Schrodinger picture field and its conjugate momentum. $\lambda_0$ and $m_0$ are the bare coupling and bare meson mass.

Cahill, Comtets and Glauber long ago derived the one-loop (order $O(\lambda_0^0)$) kink mass in this model \cite{cahill76}
\beq
Q=Q_0+Q_1\hsp Q_0=\frac{m_0^3}{3\lambda_0}\hsp Q_1=\left(\frac{1}{4\sqrt{3}}-\frac{3}{2\pi}\right)m_0.
\eeq

\section{Arbitrary Redefinitions} \label{arbsez}

Let us define two parameters $\lambda$ and $m$ and also the differences
\beq
\delta\sl=\sl-\sqrt{\lambda_0}\hsp \delta m^2=m^2-m_0^2.
\eeq
Consider the dimensionless ratio $\lambda/m^2$ small and positive.  Then we may expand in powers of $\lambda/m^2$ 
\beq
\delta\sl=\sum_i \delta\sl_{i}\hsp \delta m^2=\sum_i \delta m^2_i
\eeq
where we define $\delta\sl_i$ and $\delta m^2_i$ to be the terms in the expansion of order $O(\lambda^i)$.

We will demand that the leading terms are $\delta\sl_{3/2}$ and $\delta m^2_1$.  To avoid clutter, we will then drop the subscripts.  Then, up to order $O(\lambda^0)$, we may rewrite the kink mass as
\beq
Q=Q_0+Q_1+O(\lambda/m^2)\hsp Q_0=\frac{m_0^3}{3\lambda_0}\hsp Q_1=\left(\frac{1}{4\sqrt{3}}-\frac{3}{2\pi}\right)m.
\eeq
Here $Q_1$ is the same quantity as appeared in Ref.~\cite{dhn2}, although the mass renormalization there was different.

We are interested in the quantity
\beq
\Delta Q= Q-\frac{m^3}{3\lambda}.
\eeq
Note that while $Q$ is an observable mass, $\Delta Q$ depends on our choice of parameters $m$ and $\lambda$ which later will depend on our choice of renormalization scheme.

Using our expansion, we find, up to corrections of order $O(\lambda)$ 
\bea
\Delta Q&=&\frac{m_0^3}{3\lambda_0}-\frac{m^3}{3\lambda}+Q_1=\frac{m^3}{3\lambda}\left(1-\frac{\delta m^2}{m^2}\right)^{3/2}\left(1-\frac{\delta\sl}{\sl}\right)^{-2}-\frac{m^3}{3\lambda}+Q_1\nonumber\\
&=&\frac{m^3}{\lambda}\left(\frac{2\delta \sl}{3\sl}-\frac{\delta m^2}{2m^2}\right)+\left(\frac{1}{4\sqrt{3}}-\frac{3}{2\pi}\right)m. \label{pad}
\eea
This is our master formula for $\Delta Q$, valid for any renormalization scheme.

\section{Renormalization} \label{rensez}

\subsection{Coupling Constant Renormalization}

Following Eq.~(34) of Ref.~\cite{gw}, the amputed three point function is
\beq
\Gamma_3(0,0)=\Gamma_{3a}+\Gamma_{3b}+\Gamma_{3c}.
\eeq
Our notation is related that of Ref.~\cite{gw} by
\beq
\lambda=\frac{\lambda_{GW}}{2}\hsp m=\mu_{GW}
\eeq
where the subscript $GW$ refers to Ref.~\cite{gw}.
Then the first term is
\beq
\Gamma_{3a}=\frac{9}{4}\frac{m\  2^{3/2}\lambda^{3/2}}{4\pi}\int_0^1 d\alpha [m^2]^{-1}=\frac{9\lambda^{3/2}}{4\sqrt{2}\pi m}.
\eeq
The second is
\beq
\Gamma_{3b}=-\frac{9im^3}{2}2\sqrt{2}\lambda^{3/2}\int\frac{d^2l}{(2\pi)^2}\frac{1}{(l^2-m^2+i\epsilon)^3}=-9\sqrt{2}
im^3\lambda^{3/2}\left(-\frac{i}{8\pi m^4} \right)=-\frac{9\lambda^{3/2}}{4\sqrt{2}\pi m}.
\eeq
The third arises from the counterterm $c_1$, to be defined below, and is
\beq
\Gamma_{3c}=\frac{c_1}{2}\frac{m}{\sqrt{2\lambda}}. \label{c1}
\eeq

In Eq.~(35) of Ref.~\cite{gw} the renormalization condition $\Gamma_3(0,0)=0$ is imposed and so we find
\beq
c_1=0. \label{zeq}
\eeq

\subsection{Mass Renormalization}

In Ref.~\cite{gw} the mass renormalization involves two diagrams, one arising from a single interaction and one from two interactions.  The single interaction diagram vanishes in the case at hand because our Hamiltonian~(\ref{heq}) is normal ordered.   The remaining polarization function consists of two terms
\beq
\Pi=\Pi_1+\Pi_2.
\eeq
The first arises from a two-vertex loop diagram and is
\beq
\Pi_1=\frac{9m^2\ 2\lambda}{16\pi}\int_0^1 d\alpha \frac{1}{m^2(1-\alpha(1-\alpha))}=\frac{9\lambda}{8\pi}\frac{2\pi}{3\sqrt{3}}=\frac{\sqrt{3}\lambda}{4}.
\eeq
The second is the counterterm
\beq
\Pi_2=\frac{c_2}{2}+\frac{m^2}{4\lambda}c_1. \label{p2}
\eeq

Using (\ref{zeq}), the renormalization condition $\Pi=0$ leads to
\beq
c_2=-\frac{\sqrt{3}}{2}\lambda. \label{c2v}
\eeq

\section{The Kink Mass and Domain Wall Tension} \label{tensez}

We have determined $\Delta Q$ as a function of our choices $\delta m^2$ and $\delta\sl$.  We also evaluated $c_1$ and $c_2$, which at this point are simply expressions for loop corrections.  The choice of renormalization condition is a choice of identification of the loop corrections $c_1$ and $c_2$ with the counterterm coefficients $\delta m^2$ and $\delta\sl$.  We will now consider two such choices.

\subsection{Scheme A} \label{asez}

The classical ground states of $H$ are $\phi=\pm m_0/\sqrt{2\lambda_0}$.  Let us decompose the Schrodinger picture quantum field about one of the ground states
\beq
\phi(x)=\frac{m_0}{\sqrt{2\lambda_0}}+\eta(x).
\eeq
Our Hamiltonian can then be written in terms of $\eta(x)$
\beq
H=\int dx\left[\frac{:\pi^2(x):+:\left(\partial_x\eta\right)^2:}{2}+\frac{\lambda_0}{4}:\eta^2\left(\eta+m_0\sqrt\frac{2}{\lambda_0}\right)^2:\right]. \label{heq2}
\eeq
The cubic interaction corresponds to the term
\beq
H_3=m_0\sqrt{\frac{\lambda_0}{2}}\int dx :\eta^3:=m\sqrt{\frac{\lambda}{2}}\left(1-\frac{\delta m^2}{2m^2}-\frac{\delta\sl}{\sl} 
\right)\int dx :\eta^3:.
\eeq

Eq.~(\ref{c1}) is the amputated three-point function for $\eta$ if we define the counterterms using this $\eta$ decomposition.  In other words, we will define $c_1$ by
\beq
H_3=m\sqrt{\frac{\lambda}{2}}\int dx :\eta^3:-\Gamma_{3c}\int dx :\eta^3:
\eeq
so that
\beq
c_1=\lambda\left(\frac{\delta m^2}{m^2}+2\frac{\delta\sl}{\sl}\right).
\eeq

Similarly, the mass term in Eq.~(\ref{heq2}) is
\beq
H_2=\frac{m_0^2}{2}\int dx :\eta^2:=\frac{m^2}{2}\int dx :\eta^2:-\frac{\delta m^2}{2}\int dx :\eta^2:.
\eeq
Identifying
\beq
H_2=\frac{m^2}{2}\int dx :\eta^2:-\Pi_2 \int dx :\eta^2:
\eeq
we obtain
\beq
c_2+\frac{m^2}{2\lambda}c_1=\delta m^2.
\eeq
Now our results (\ref{zeq}) and (\ref{c2v}) lead to 
\beq
\frac{\delta m^2}{m^2}=-\frac{\sqrt{3}}{2}\frac{\lambda}{m^2}\hsp \frac{\delta\sl}{\sl}=\frac{\sqrt{3}}{4}\frac{\lambda}{m^2}.
\eeq
Substituting these into our master formula (\ref{pad}) we obtain
\beq
\Delta Q=\left(\frac{\sqrt{3}}{2}-\frac{3}{2\pi}\right)m
\eeq
in agreement with Ref.~\cite{noifin}.

\subsection{Scheme B} \label{bsez}

The counterterms corresponding to the parameters $c_1$ and $c_2$ are given in Eq.~(31) of Ref.~\cite{gw}
\beq
H_{ct}=\int dx \left[-\frac{c_1}{8}:\left(\phi^2-v^2\right)^2:-\frac{c_2}{2}:\left(\phi^2-v^2\right):\right].
\eeq
We have included normal ordering, which was not used in Ref.~\cite{gw}.
The definition of $v$ is somewhat subtle, as the expectation value of $\phi$ on a state depends on the state and the renormalization conditions.  However, the fact that $c_2$ does not appear in the amputated three-point function suggests that
\beq
v=\frac{m}{\sqrt{2\lambda}}.
\eeq
We will assume this.  Any other choice does not affect the kink mass, but does affect our renormalization conditions by changing the definition of the three-point coupling.  This is because the three-point coupling is defined to be the third derivative evaluated at a certain value of the classical field, and so it is necessarily sensitive to this choice.  Therefore a change in $v$ will change our renormalization scheme and so will change $\delta\sl$.

With this caveat, we write a Hamiltonian as
\beq
H=H_{ct}+\int{dx}\left[\frac{:\pi^2:+:\left(\partial_i\phi\right)^2:}{2}+\frac{\lambda}{4}:\left(\phi^2-\frac{m^2}{2\lambda}\right)^2:\right]. 
\eeq
The potential terms in the classical Hamiltonian density are
\beq
U(\phi)=\frac{\lambda-\frac{c_1}{2}}{4}\left(\phi^2-\frac{m^2}{2\lambda}\right)^2-\frac{c_2}{2}\left(\phi^2-\frac{m^2}{2\lambda}\right).
\eeq
This of course is equal to the original potential $(\lambda_0/4) (\phi^2-m_0^2/(2\lambda_0))^2$ and so
\beq
c_1=4\sl\delta\sl\hsp c_2=m^2\left(\frac{\delta\sl}{\sl}-\frac{\delta m^2}{2m^2}\right).
\eeq

Now the loop corrections 
(\ref{zeq}) and (\ref{c2v}) lead to 
\beq
\delta\sl=0\hsp \delta m^2=\sqrt{3}\lambda
\eeq
and so
\beq
\Delta Q=\left(-\frac{5\sqrt{3}}{12}-\frac{3}{2\pi}\right)m\sim -1.199153m
\eeq
consistent with the result of Ref.~\cite{gw}.

\subsection{Domain Walls}

As shown in Ref.~\cite{noi3d}, the Cahill, Comtets and Glauber formula applies to the domain wall tension $Q$ in any dimension
\beq
Q=Q_0+Q_1\hsp Q_0=\frac{m_0^3}{3\lambda_0}.
\eeq
In the case of the $\phi^4$ double-well model, the formula for the classical tension $Q_0$ is independent of the dimension.  On the other hand, the one-loop correction $Q_1$ does depend on the dimension.  However, critically, it is independent of the scheme.  As a result, in any dimension our master formula may still be written
\beq
\Delta Q = \frac{m^3}{\lambda}\left(\frac{2\delta \sl}{3\sl}-\frac{\delta m^2}{2m^2}\right)+Q_1.
\eeq

Let us use subscripts $A$ and $B$ to respectively denote the schemes in Subsecs.~\ref{asez} and \ref{bsez} respectively.  The general arguments above show that
\bea
c_1&=&\lambda\left(\frac{\delta_A m^2}{m^2}+2\frac{\delta_A\sqrt{\lambda}}{\sl}\right)=\lambda\left(4\frac{\delta_B\sqrt{\lambda}}{\sl}\right)\\\
c_2&=&m^2\left(\frac{\delta_A m^2}{2m^2}-\frac{\delta_A\sqrt{\lambda}}{\sl}\right)=m^2\left(-\frac{\delta_B m^2}{2m^2}+\frac{\delta_B\sqrt{\lambda}}{\sl}\right).
\eea
Therefore the counterterm coefficients are related by
\beq
\frac{\delta_B\sl}{\sl}=\frac{\delta_A m^2}{4m^2}+\frac{\delta_A\sl}{2\sl}\hsp
\frac{\delta_Bm^2}{m^2}=-\frac{\delta_A m^2}{2m^2}+\frac{3\delta_A\sl}{\sl}.
\eeq
The resulting difference in the two loop tension corrections is
\beq
\Delta Q_B-\Delta Q_A=\frac{m^3}{\lambda}\left(\frac{2\delta_B \sl}{3\sl}-\frac{\delta_B m^2}{2m^2}
-
\frac{2\delta_A \sl}{3\sl}+\frac{\delta_A m^2}{2m^2}
\right)=\frac{11m^3}{12\lambda}\left(\frac{\delta_A m^2}{m^2}-\frac{2\delta_A \sl}{\sl}
\right).
\eeq
In the case of the domain wall string in 2+1 dimensions \cite{noifin}
\beq
\frac{\delta_A m^2}{m^2}=-\frac{9{\rm{ln}}(3)}{8\pi}\frac{\lambda}{m}\hsp \frac{\delta_A\sl}{\sl}=\frac{9({\rm{ln}}(3)-1)}{16\pi}\frac{\lambda}{m}
\eeq
and so
\beq
\Delta Q_B-\Delta Q_A=\frac{33}{32\pi}(1-2{\rm{ln}}(3))m^2\sim -0.392997m^2
\eeq
which is indeed the difference between the two results shown on Table IV of Ref.~\cite{gw}.

In the case of 3+1 dimensions \cite{noifin}
\bea
\frac{\delta_A m^2}{m^2}&=&-\frac{9\lambda}{2\pi^2}\int_0^\infty dp \frac{p^2}{\omega_p(3m^2+4p^2)}\\ \frac{\delta_A\sl}{\sl}&=&\frac{9\lambda}{16\pi^2}+\frac{9\lambda}{8\pi^2}\int_0^\infty dp \ p^2\left(-\frac{1}{\omega_p^3}+\frac{2}{\omega_p(3m^2+4p^2)}
\right)\nonumber
\eea
and so
\bea
\Delta Q_B-\Delta Q_A&=&-\frac{33m^{3}}{32\pi^2}-\frac{33m^{3}}{4\pi^2}\int_0^\infty dp \ p^2\left(-\frac{1}{4\omega_p^3}+\frac{1}{\omega_p(3m^2+4p^2)}
\right)\\
&=& \left(-\frac{99}{32\pi^2}+\frac{11\sqrt{3}}{32\pi}\right)m^3\sim -0.123943m^3\nonumber
\eea
which is again consistent with Table IV of Ref.~\cite{gw}.

\section{Remarks}

We have shown that the spectral methods, with dimensional regularization, of Refs.~\cite{rebmuro,gw} yield the same domain wall tension as the Hamiltonian methods of Ref.~\cite{noifin} which employ a hard cutoff. 

There were many potential problems with both approaches.  For example, a hard cutoff in Ref.~\cite{rebhan} yielded the wrong answer.  In fact, the problem with the hard cutoff in Ref.~\cite{rebhan} is that the vacuum and kink sectors were regularized separately and then there was an {\it{ad hoc}} matching between the two sectors, whereas in Ref.~\cite{noifin} the theory was regularized only once and the two sectors were treated with the same Hamiltonian.  Similarly, one might have worried that dimensional regularization would be ill-defined in a configuration that solves the classical field equations in an integral dimension.  

However the agreement that we have found here puts both worries to rest.  At least this is true for the present case, in which the divergences are only logarithmic and the field configuration is trivial along the directions that are regularized.  It remains to be seen whether one method will have a problem in other applications without these features.  Needless to say, we believe that such a comparison between the results of different methods will be essential for progress towards a treatment of more complicated but interesting solitons, such as the 't Hooft-Polyakov monopole or the gravitating kink \cite{zhong21,zhong24}.




\section* {Acknowledgement}

\noindent
This work was supported by the Higher Education and Science Committee of the Republic of Armenia (Research Project Nos. 24RL-1C047 and 24PostDoc/2‐1C009).

\end{document}